\documentclass[twocolumn,showpacs,aps,prl,reprint,groupaddress]{revtex4-1}

\pdfoutput=1

\usepackage{epsfig}
\usepackage{amsmath}
\usepackage{amssymb}
\usepackage{amsfonts}
\usepackage{bm}
\usepackage{color}

\newcommand{\beq}{\begin{equation}}
\newcommand{\eeq}{\end{equation}}
\newcommand{\beqarray}{\begin{eqnarray}}
\newcommand{\eeqarray}{\end{eqnarray}}
\newcommand{\eq}[1]{Eq.~(\ref{#1})} 
\newcommand{\fig}[1]{Fig.~(\ref{#1})} 
\newcommand{\Ref}[1]{Ref.~\onlinecite{#1}} 

\newcommand{\sos}[2]{\ensuremath{\hat{s}^{#1}\otimes\hat{\sigma}^{#2}}}

\begin{document}

\allowdisplaybreaks

\title{Odd-parity superconductivity from phonon-mediated
    pairing: Application to Cu$_x$Bi$_2$Se$_3$}
\author{P. M. R. Brydon}
\email{pbrydon@umd.edu}
\affiliation{Condensed Matter Theory Center and Joint Quantum
  Institute, University of Maryland, College Park, Maryland
  20742-4111, USA}
\author{S. Das Sarma}
\affiliation{Condensed Matter Theory Center and Joint Quantum
  Institute, University of Maryland, College Park, Maryland
  20742-4111, USA}
\author{Hoi-Yin Hui}
\affiliation{Condensed Matter Theory Center and Joint Quantum
  Institute, University of Maryland, College Park, Maryland
  20742-4111, USA}
\author{Jay D. Sau}
\affiliation{Condensed Matter Theory Center and Joint Quantum
  Institute, University of Maryland, College Park, Maryland
  20742-4111, USA}

\date{\today}

\begin{abstract}
Motivated by the proposed topological state in Cu$_x$Bi$_2$Se$_3$, we
study the possibility of phonon-mediated odd-parity superconductivity
in spin-orbit coupled systems with time-reversal and inversion
symmetry. For such systems, we show that, in general, pure
electron-phonon coupling can never lead to a triplet state with a
higher critical temperature than the leading singlet state. The 
Coulomb pseudopotential,  which is the repulsive part of the 
electron-electron interaction and is typically small in weakly
correlated systems, is therefore critical to stabilizing the
triplet state. We introduce a chirality quantum number, which
identifies the electron-phonon vertex interactions that are most
favorable to the triplet channel as those that conserve
chirality. Applying these results to Cu$_x$Bi$_2$Se$_3$, we find that 
a phonon-mediated odd-parity state may be realized in the presence of
weak electronic correlations if the chirality-preserving
electron-phonon vertices are much stronger than the chirality-flipping 
vertices.
\end{abstract}


\maketitle

{\it Introduction.}---The discovery that gapped electronic systems
can be topologically nontrivial has sparked enormous
interest~\cite{Schnyder2008,HasanKane2010}. While there now exists several clear
examples of topological insulators, such as Bi$_2$Se$_3$~\cite{Bi2Se3}
and SnTe~\cite{SnTe}, the  
unconventional gap structure of topological superconductors make
these systems much rarer~\cite{Qi2010}. Intriguingly, a superconducting state
appears upon doping some topological insulators, most notably 
Cu$_x$Bi$_2$Se$_3$~\cite{Hor2010}. Fu and Berg have proposed that this
system realizes a topological superconductor, with a novel odd-parity
(triplet) pairing state~\cite{FuBerg2010}. 

Cu$_x$Bi$_2$Se$_3$ has subsequently been the subject of intense
study~\cite{Kriener2011,Bay2012,subgapstates,nosubgapstates,Yip2013,boundstatestheory,Das2011,WanSavrasov2014,Hasan2010}. 
Experiments show a full gap~\cite{Kriener2011,Bay2012}, and
anomalies in the dc magnetization~\cite{Das2011} and an upper
critical field that exceeds the Pauli limit indicate triplet
pairing~\cite{Bay2012}. This interpretation is supported by 
point-contact spectroscopy measurements of the expected
topologically-protected surface subgap
states~\cite{subgapstates,boundstatestheory,Yip2013}, but other
experiments find no subgap structure~\cite{nosubgapstates}, consistent
with nontopological $s$-wave pairing.  
Although the experimental situation in Cu$_x$Bi$_2$Se$_3$ has not yet been
settled, similar
signatures of unconventional superconductivity have been observed 
in Sn$_{1-x}$In$_{x}$Te~\cite{Sasaki2012} and Bi$_2$Se$_3$ under
pressure~\cite{Kirschenbaum2013}. This raises the tantalizing
possibility of an entire class of topological superconductors obtained
by doping topological insulators. 

The origin of a triplet pairing state in any of these doped
semiconductors is mysterious, as they are likely free of
the strong correlations thought to be an essential aspect~\cite{Anderson1984}
of the triplet superconductors
UPt$_3$~\cite{UPt3} and Sr$_2$RuO$_4$~\cite{Sr2RuO4}. Rather, 
the electron-phonon interaction is expected to play the dominant role in
the pairing~\cite{FuBerg2010,WanSavrasov2014,Sasaki2012}. This is quite
surprising, however, as it is widely believed that phonon-mediated
pairing generically yields a singlet
state~\cite{phonon_triplet}, although a definitive proof
has been lacking. Furthermore, previous analyses did not
include the strong spin-orbit coupling characteristic
of topological insulators and which may favor triplet
pairing~\cite{FuBerg2010,Sasaki2012}. As such, they cannot exclude the
possibility that the electron-phonon interaction indeed stabilizes a
triplet state in these materials.

In this paper we study the fundamental question of when 
electron-phonon interactions stabilize a triplet state,
 and thus evaluate the conditions required for the proposed topological 
superconductivity in Cu$_x$Bi$_2$Se$_3$.
We first prove a theorem, showing that for the BCS theory the
symmetries of the electron-phonon vertex  
functions ensure that, purely with electron-phonon coupling, the
critical temperature of the leading triplet state {\it never} exceeds
that of the leading singlet. Therefore, the stabilization of the triplet
state must depend on the so-called Coulomb pseudopotential, which
may not be small~\cite{Bauer2013}. We then define a generalized
chirality operator, 
which allows us to identify electron-phonon coupling vertices that
would stabilize a triplet gap. Materials where chirality preserving
vertices dominate could be candidates for electron-phonon mediated
triplet superconductivity. Finally, we apply these
insights to a model of Cu$_x$Bi$_2$Se$_3$~\cite{FuBerg2010}, and
identify the electron-phonon vertices that cause an attractive
interaction in the triplet channel. If these terms dominate the
electron-phonon interaction, the topological state could be
realized in the presence of weak correlations. 

{\it Electron-phonon interaction and pairing.}---We start by
considering a strongly spin-orbit coupled system with 
inversion (${\cal I}$) and time-reversal (${\cal T}$) 
symmetries, so that every eigenstate is at least doubly
degenerate~\cite{NCS}. Assuming for simplicity that a single band
crosses the Fermi energy, we can index the
degenerate states by a pseudospin variable $s=\pm$, such that ${\cal I}|{\bf
  k},s\rangle = |-{\bf k},s\rangle$ and ${\cal T}|{\bf
  k},s\rangle = s|-{\bf k},-s\rangle$. In the presence of strong
spin-orbit coupling the electron-phonon interaction may not conserve
pseudospin (in contrast to~\Ref{phonon_triplet}), and so we
have the general form  
\beq
H_{\text{e-p}} = \sum_{{\bf k},{\bf k}'}\sum_{s,s'}\sum_\eta
g^{\eta}_{s',s}({\bf k}',{\bf 
  k})(b^\dagger_{{\bf k}-{\bf k}',\eta} + b^{}_{{\bf k}'-{\bf k},\eta})c^\dagger_{{\bf k}',s'}c^{}_{{\bf k},s}\,,
\eeq
where $b_{{\bf q},\eta}$ is the annihilation operator for a phonon
in mode $\eta$ with momentum ${\bf q}$, and $c_{{\bf k},s}$ is the
annihilation operator for an electron in state $|{\bf k},s\rangle$.
The inversion and time-reversal symmetries
require that the vertex functions satisfy $g^{\eta}_{s',s}({\bf k}',{\bf  
  k}) = \pm_{\eta} g^{\eta}_{s',s}(-{\bf k}',-{\bf 
  k})$ and $g^{\eta}_{s',s}({\bf k}',{\bf k}) = ss'[g^\eta_{-s',-s}(-{\bf
    k}',-{\bf k})]^\ast$, respectively, where the sign $\pm_\eta$
under inversion depends on the phonon mode.

Within the BCS approximation, the electron-phonon coupling generates
the pairing interaction 
\beqarray
V_{s_2,s_1;s_3,s_4}({\bf k}',{\bf k}) &=& -\sum_\eta
\frac{g^\eta_{s_1,s_3}({\bf k}',{\bf k})g^\eta_{s_2,s_4}(-{\bf
  k}',-{\bf k})}{\omega_{{\bf k}'-{\bf k},\eta}} \notag \\ 
&& \times \Theta(\omega_D-|\epsilon_{{\bf
    k}}|)\Theta(\omega_D-|\epsilon_{{\bf k}'}|)\,, \label{eq:pairing}
\eeqarray
where $\omega_{{\bf q},\eta}$ is the dispersion of phonon mode
$\eta$, $\epsilon_{\bf k}$ is the electronic dispersion, and
$\omega_D$ is a cutoff on the order of the Debye 
energy. The pairing interaction is the kernel of the linearized
BCS equation for the matrix gap function $\hat{\Delta}({\bf k})$, which is
formulated as an eigenvalue problem 
\beq
\lambda \Delta_{s_1,s_2}({\bf k}') = -\sum_{{\bf
    k},s_3,s_4}V_{s_2,s_1;s_3,s_4}({\bf k}',{\bf k})
\Delta_{s_3,s_4}({\bf k})\,. \label{eq:linearBCS}
\eeq
Only solutions with positive eigenvalues have a finite critical
temperature, and the solution with the largest eigenvalue is the leading
instability. Inversion symmetry limits physical
solutions to either even-parity
pseudospin singlet or odd-parity pseudospin triplet states. 

{\it Singlet vs. triplet pairing.}---In the conventional case, i.e. in
the absence of spin-orbit coupling, electron-phonon  
coupling is expected to lead to the singlet channel being
dominant. Such a singlet pairing state is described by a gap
function $\hat{\Delta}^{(s)}({\bf k})=f^{(s)}({\bf
  k})(i\hat{\sigma}^y)$, where $f^{(s)}({\bf k})$ gives the momentum 
dependence of the pairing function. For the general electron-phonon
interaction, the symmetries of the electron-phonon vertices yield a gap 
equation in the singlet channel of the form  
\beq
\lambda^{(s)} f^{(s)}({\bf k}') = \sum_{{\bf
    k},s,\eta}\frac{|g^\eta_{ss}({\bf k}',{\bf k})|^2+|g^\eta_{s\bar{s}}({\bf k}',{\bf k})|^2}{\omega_{{\bf k}-{\bf k}',\eta}}
f^{(s)}({\bf k})\,, \label{eq:linearBCSsing}
\eeq
where the momenta are restricted to the shell of thickness
$\omega_D$ about the Fermi surface.
The singlet gap function is therefore an eigenstate of a matrix with
nonnegative entries. It follows from the Perron-Frobenius
theorem~\cite{PF} that the gap function $f^{(s)}({\bf k})$ of the
dominant instability has no sign changes as a function of the
wavevector ${\bf k}$, as is characteristic of conventional singlet
pairing.  

We now consider the triplet pairing function with the
largest critical temperature, 
$\hat{\Delta}^{(t)}({\bf k})$. To compare with the singlet
channel, we apply a momentum 
dependent pseudospin-rotation transformation so that it is recast in
the form $\hat{\Delta}^{(t)}({\bf k}) = 
\chi_{\bf k}f^{(t)}({\bf k})\hat{\sigma}^x$, where
$f^{(t)}({\bf k})$ and $\chi_{\bf k}$ are the magnitude and sign
of the triplet gap, respectively. In other words, we have
rotated  the pseudospin at ${\bf k}$ and $-{\bf k}$ so that in the new 
pseudospin basis the triplet pair formed from these states has
vanishing $z$-component of pseudospin. Note that this rotation  does
not affect the singlet pairing, nor does it alter the symmetry properties of
the electron-phonon vertices. 
The gap magnitude $f^{(t)}({\bf k})$ satisfies the eigenvalue
equation
\beq
\lambda^{(t)} f^{(t)}({\bf k}') = \sum_{{\bf
    k},s,\eta}\chi_{{\bf k}'}\chi_{{\bf k}}\frac{|g^\eta_{ss}({\bf k}',{\bf k})|^2-|g^\eta_{s\bar{s}}({\bf k}',{\bf k})|^2}{\omega_{{\bf k}-{\bf k}',\eta}}
f^{(t)}({\bf k})\,. \label{eq:linearBCStrip}
\eeq
The magnitude of the matrix elements in~\eq{eq:linearBCStrip} are
bounded by the corresponding elements in the singlet gap
equation. By a corollary to the 
Perron-Frobenius theorem~\cite{PF}, the maximal eigenvalue
of~\eq{eq:linearBCStrip} therefore cannot exceed the maximal singlet
eigenvalue. Since the leading triplet gap 
satisfies~\eq{eq:linearBCStrip}, we have our first major result
which can be stated as the following theorem: in a 
system with inversion and time-reversal symmetry, the critical 
temperature of the leading triplet gap {\it never exceeds} that of 
the leading singlet gap for a purely phonon-mediated pairing
interaction. 

Our analysis implies that electronic correlations are vital
to stabilizing a triplet state. In particular, the spatial
separation of the electrons in a triplet Cooper pair reduces the
pair-breaking effect of the Coulomb pseudopotential compared to a
$s$-wave singlet state. A sufficiently large Coulomb pseudopotential
may therefore reduce the critical
temperature of the leading singlet state below that of the
triplet~\cite{phonon_triplet}. Such a strong Coulomb
pseudopotential is the necessary condition for the triplet
superconductivity to emerge in the system.

{\it Degenerate singlet and triplet states}.---While the singlet
pairing typically may be expected to dominate over 
triplet pairing, it was pointed out by Fu and
Berg~\cite{FuBerg2010} that the singlet and triplet states would be
degenerate if the Dirac-like Hamiltonian considered by them commuted
with a chirality operator. Motivated by this, we
generalize the notion of ``chirality'' to index the doubly-degenerate
states near the Fermi surface of an arbitrary electronic
system. Specifically, the pseudospin states $|{\bf
  k},s\rangle$ become the chiral states $|{\bf
  k},\nu\rangle$ where the chirality $\nu = s\chi_{\bf k}$, and
$\chi_{\bf k}$ is the sign of the leading triplet gap as 
defined above. We hence replace the pseudospin indices in the gap
equations~\ref{eq:linearBCSsing} and~\ref{eq:linearBCStrip}  by
chirality indices using $g^\eta_{s',s}({\bf k}',{\bf k}) =
g^\eta_{\nu',\nu}({\bf k}',{\bf k})\delta_{\nu',s'\chi_{{\bf
      k}'}}\delta_{\nu,s\chi_{\bf k}}$, obtaining
\beq
\lambda^{(\alpha)} f^{(\alpha)}({\bf k}') = \sum_{{\bf
    k},\nu,\eta}\frac{|g^\eta_{\nu\nu}({\bf k}',{\bf k})|^2\pm|g^\eta_{\nu\bar{\nu}}({\bf k}',{\bf k})|^2}{\omega_{{\bf k}-{\bf k}',\eta}}
f^{(\alpha)}({\bf k}),
\eeq 
where the plus (minus) sign in the summand holds for $\alpha=s$ ($t$).
Comparing the transformed equations in the singlet and triplet
channels, it is clear that the singlet and triplet eigenvalues
are identical if
the electron-phonon vertices do not flip the chirality index,
i.e. $\lambda^{(s)}=\lambda^{(t)}$. We see that
electron-phonon vertices which preserve an appropriately-defined
chirality index generate
attractive interactions in the triplet channel, while
chirality-flipping vertices are always triplet pair-breaking. This is the
second major result of our paper. Note that in previous works, unconventional
pairing is achieved via a strongly forward-scattering
electron-phonon interaction, which promotes attractive interactions
in many pairing channels~\cite{WanSavrasov2014,forward}. In
contrast, our condition precisely determines the 
electron-phonon interactions that generate the triplet state, and
there is no requirement that these 
vertices involve small momentum transfers.

We make our discussion more concrete by using the
chirality index to define a chirality
operator ${\cal O}_{\text{ch}}({\bf k}) = \sum_\nu\nu|{\bf
  k},\nu\rangle\langle {\bf k},\nu|$. When only electron-phonon   
interactions which commute with $\sum_{\bf k}{\cal O}_{\text{ch}}({\bf
  k})$ are present, every singlet solution $\hat{\Delta}^{(s)}({\bf 
  k})$ is degenerate with a triplet solution $\hat{\Delta}^{(t)}({\bf
  k}) = U({\bf k})\hat{\Delta}^{(s)}({\bf
  k})U(-{\bf k})$, where $U({\bf k}) = \exp(i\pi{\cal
  O}_{\text{ch}}({\bf k})/4)$. On the other 
hand, an electron-phonon interaction which does not commute with the
chirality operator is triplet pair-breaking. Crucially,
it is not necessary to solve the gap equations
to define the chirality index, as this only depends upon the sign
structure of the triplet gap. This is very convenient, as it is
common to approximate the exact solution of the gap equations by a
simple function consistent with the point group. Given such a
time-reversal-invariant triplet state, we can hence define a
chirality operator which relates it to a singlet state with
nonnegative gap. The
effective coupling constants for these two states, obtained by
taking the inner product of the gap functions with the pairing
interaction~\eq{eq:pairing}, are then degenerate if only 
electron-phonon vertices which preserve the chirality are present.

{\it Application to Cu$_x$Bi$_2$Se$_3$.}---The proposed odd-parity
pairing state of Cu$_x$Bi$_2$Se$_3$ provides an excellent illustration
of the preceding discussion. We start by introducing an effective
Hamiltonian valid near the Fermi surface, where the electronic states are 
primarily derived  from the Se $p_z$-orbitals at the top and bottom of
the quintuple-layer unit cell. Denoting these two distinct sites by
$s^{z}=\pm1$, the low-energy spectrum is described
by the ${\bf k}\cdot{\bf p}$ model~\cite{FuBerg2010}
\beqarray
H_{0} &=& \sum_{\bf k}\psi^\dagger({\bf k})\left[-\mu\sos{0}{0} +
  m\sos{x}{0} + v_zk_z\sos{y}{0}\right. 
\notag \\
&& \left. + v\left(k_x\sos{z}{y} -
k_y\sos{z}{x}\right)\right]\psi^{}({\bf k})\,. \label{eq:kpHam}
\eeqarray
Here $\psi^{}({\bf k}) = (c^{}_{{\bf k},1,\uparrow},c^{}_{{\bf
    k},1,\downarrow},c^{}_{{\bf k},-1,\uparrow},c^{}_{{\bf
    k},-1,\downarrow})^T$, where $c^{}_{{\bf k},n,\sigma}$ destroys an
electron with momentum ${\bf k}$ and spin $\sigma$ at site~$n$. The
Pauli matrices in site and spin space are denoted by $\hat{s}^\mu$ and
$\hat{\sigma}^\mu$, respectively. The chemical potential is denoted by
$μ$, $m$ is the mass, and $v_z$ and $v$ are velocities along the $z$-axis
and in the $x$-$y$ plane, respectively.
We consider the physical case where
the chemical potential lies in the conduction band, i.e. $\mu>m$. 
The Hamiltonian is symmetric under  inversion (${\cal
  I}=\sos{x}{0}$) and time-reversal (${\cal T} = i\sos{0}{y}{\cal
  K}$), and so the eigenstates of~\eq{eq:kpHam}
can be labeled by a pseudospin~\cite{Yip2013}. 

The site degree of freedom allows odd-parity 
superconducting states in a relative $s$-wave, such as the $A_{1u}$
state $\Delta_{A_{1u}}i\sos{y}{x}$ proposed 
in~\Ref{FuBerg2010}. As it opens a full gap 
on the Fermi surface~\cite{FuBerg2010,Yip2013}, and has surface bound states
consistent with 
point-contact spectroscopy
measurements~\cite{subgapstates,boundstatestheory,Yip2013}, it 
is one of the most promising candidates for a 
topological state in Cu$_x$Bi$_2$Se$_3$. We have seen, however, that
the phonon-mediated pairing 
interaction generally 
favors an even-parity state with a full gap. The simplest
example of this is the topologically-trivial $A_{1g}$ state 
$\Delta_{A_{1g}}i\sos{0}{y} + \Delta_{A_{1g}}^\prime
i\sos{x}{y}$~\cite{FuBerg2010}. 

In the absence of the mass term in~\eq{eq:kpHam}, the Bogoliubov
Hamiltonian for the $A_{1g}$ state with $\Delta_{A_{1g}}^\prime=0$
can be mapped into that for the $A_{1u}$  state by the unitary
transformation  
$U=\exp(i\pi\sos{y}{z}/4)$~\cite{FuBerg2010}. This immediately
identifies the chirality operator as ${\cal O}_{\text{ch}}({\bf k}) =
\sos{y}{z}$. In the compact notation of~\eq{eq:kpHam}, we have the
  general electron-phonon interaction Hamiltonian 
\beqarray
H_{\text{e-p}} & = & \sum_{{\bf k},{\bf k}'}\sum_\eta\sum_{\mu,\nu}
f^{\eta}_{\mu,\nu}({\bf 
  k}',{\bf k})\left(b^{\dagger}_{{\bf
    k}-{\bf k}',\eta} + b_{{\bf k}'-{\bf k},\eta}\right) \notag \\
&& \times\psi^{\dagger}({\bf k}')\sos{\mu}{\nu}\psi({\bf k})\,. \label{eq:HelphCuxBi2Se3}
\eeqarray
If only vertex functions $f^{\eta}_{\mu,\nu}({\bf k}',{\bf k})$ for which
$\sos{\mu}{\nu}$ commutes with the chirality operator $\sos{y}{z}$ are
nonzero, it follows from the discussion above
that the coupling constants for the $A_{1u}$ and $A_{1g}$
states are identical. Vertex
functions for which $\sos{\mu}{\nu}$ anticommutes with $\sos{y}{z}$
are generally expected to be present, however, giving the $A_{1g}$ state
the higher coupling constant.

In the general case of a finite mass gap, the Fu and Berg $A_{1g}$ and
$A_{1u}$ Hamiltonians cannot be mapped into one another by a
chirality transformation. In the vicinity of 
the Fermi surface, however, we can define a chirality
operator that relates the two gaps~\cite{pseudochiral}. This is
sufficiently close to the chirality operator in the massless limit
that the classification of the electron-phonon vertices obtained above
remains valid to good approximation. Specifically, the
chirality-preserving electron-phonon vertices for the massless case
are now either still chirality-preserving,
or contain chirality-flipping terms which are
smaller by a factor of $m/\mu\approx0.3$ than the
chirality-preserving~\cite{Hasan2010}. A 
similar analysis holds for the vertices which flip the chirality in
the $m=0$ limit. Our classification of the electron-phonon
  vertices is the starting point for a detailed microscopic
  analysis of the pairing instability in Cu$_x$Bi$_2$Se$_3$.

We make this concrete by considering a toy model where
the electrons couple to a dispersionless optical mode with frequency
$\omega_0$. From~\eq{eq:HelphCuxBi2Se3} we include only the 
$(\mu,\nu) = (0,0)$ and $(x,0)$ terms, representing chirality-preserving and
flipping vertices, respectively. We assume that the
corresponding vertex functions $g_0$ and $g_x$ are constant.
The Fu and Berg $A_{1g}$ and $A_{1u}$ states are then exact
eigenstates of the phonon-mediated pairing interaction, with
eigenvalues $\lambda_{A_{1g}} = (g^2_0 + g_x^2 + 2|g_xg_0|m/\mu)/\omega_0$ and
$\lambda_{A_{1u}} = (g_0^2 - g_x^2)(1 - (m/\mu)^2)/\omega_0$,
respectively. The
$A_{1g}$ state is the leading instability for
nonzero $g_x$ or $m$, while the $A_{1u}$ state only has finite
critical temperature for 
$|g_x|<|g_0|$. We also include the
on-site repulsion $H_{\text{e-e}} = U/V\sum_{\bf
  q}\sum_{s=\pm}\rho_{s,\uparrow}({\bf q})\rho_{s,\downarrow}(-{\bf
  q})$ where $\rho_{s,\sigma}({\bf q})=\sum_{\bf k}c^{\dagger}_{{\bf
    k}+{\bf q},s,\sigma}c^{}_{{\bf
    k},s,\sigma}$ and $V$ is the volume.
As the first $A_{1g}$ gap $\Delta_{A_{1g}}$
involves on-site pairing, a finite $U>0$ will 
tend to lower its critical temperature. On the other hand, the
intersite $A_{1u}$ state is unaffected by $H_{\text{e-e}}$. 

\begin{figure}
\includegraphics[width=\columnwidth]{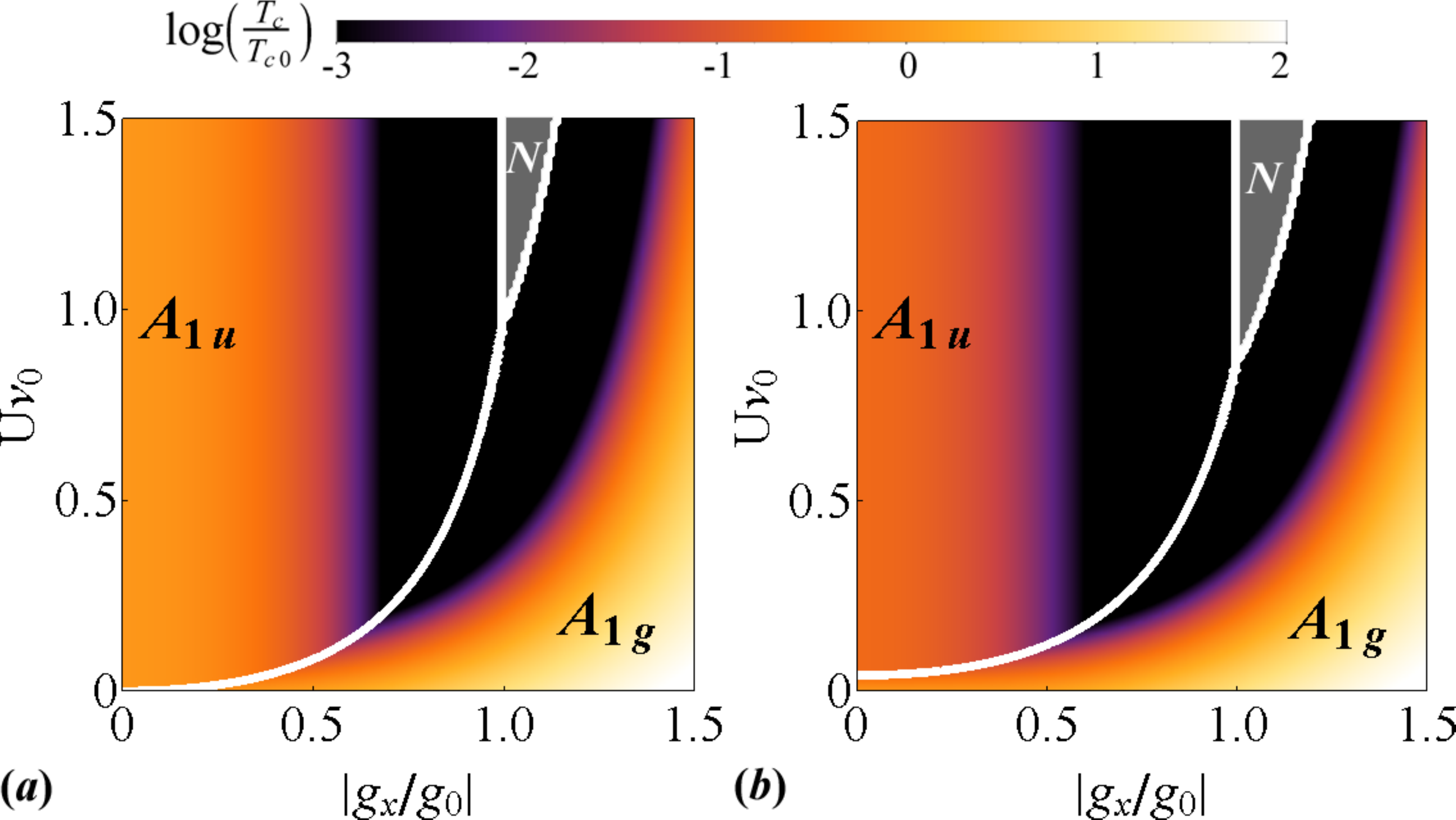}
\caption{(color online). Phase diagram for our toy model of
  Cu$_x$Bi$_2$Se$_3$ with (a)
  vanishing ($m=0$) and (b) nonzero 
  ($m=0.4\mu$) mass gap, showing the leading superconducting
  instability as a function of $|g_x/g_0|$ and $U\eta_0$. The
  logarithmic colour scale shows the critical temperature $T_c$
  relative to the critical temperature $T_{c0}$ at $m=U=g_x=0$. In
  the grey region N the system remains normal down to zero
  temperature. We set $W=10\mu$, 
  $\omega_D=0.1\mu$, and $g_0^2/\omega_0=0.1225/\eta_0$.}\label{phasediagrams}
\end{figure}

We study the pairing in our model within the mean-field
approximation. For simplicity, the conduction band is assumed to
extend from $m-\mu$ below the Fermi 
surface to $W-\mu$ above, with constant density of states
$\nu_0$ and $W\gg m$. Deriving the gap equations, we find that the
critical temperature of the $A_{1g}$ state satisfies
\beq
\det\left|\begin{array}{ccc}
\left(\frac{g_x^2+g_0^2}{\omega_0} - \frac{U}{2}\right)\chi_0 -1 &
  -\frac{U}{2}\chi &  \frac{g_0g_x}{\omega_0}\chi_{01}\\
-\frac{U}{2}\chi_0 & -\frac{U}{2}\chi -1 & 0 \\
\frac{g_0g_x}{\omega_0}\chi_{01} & 0 & \frac{g_x^2+g_0^2}{\omega_0}\chi_1 
-1\end{array}\right| = 0\,,
\eeq
while for the $A_{1u}$ state we have to solve
$\lambda_{A_{1u}}\chi_0 = 1$. Following the notation
of~\Ref{FuBerg2010}, the gap equations are expressed in terms of 
$\chi_0 = {\nu}_0\int^{\omega_D}_{-\omega_D}d\epsilon
\tanh(\epsilon/2k_bT_c)/\epsilon$, $\chi_{01} = (m/\mu)\chi_0$,
$\chi_{1}=(m/\mu)^2\chi_0$, 
and $\chi={\nu}_0\int^{W-\mu}_{m-\mu}d\epsilon 
\tanh(\epsilon/2k_bT_c)/\epsilon - \chi_0$.
The resulting phase diagram is shown in~\fig{phasediagrams} for the
cases of (a) vanishing and (b) nonzero mass gap. In the
absence of on-site repulsion the $A_{1g}$ state has higher critical
temperature than the $A_{1u}$, except for $m=g_x=0$ where the two
are degenerate. Sufficiently strong on-site repulsion
suppresses the critical temperature of the $A_{1g}$ state below that for
the $A_{1u}$. For small ratios $|g_x/g_0|\lesssim 0.5$, this requires
only a relatively weak repulsion $U\approx0.1W$. If $|g_x/g_0|$ is close
to unity, however, a repulsive potential on the order of the bandwidth
is necessary, and the critical temperature will be very small. 
Since Cu$_x$Bi$_2$Se$_3$ is likely weakly-correlated, we conclude
that the $A_{1u}$ state could be realized if the
chirality-preserving electron-phonon vertices are much larger than the
chirality-flipping, which is the final major result of our work. It is
not obvious that this should be the case, however,
and this problem requires detailed microscopic modeling beyond the
present discussion. Interestingly, Wan and Savrasov
have recently proposed that a strongly forward-scattering
phononic modulation of the spin-orbit coupling is generic to layered
semiconductors~\cite{WanSavrasov2014}, although a nodal
$A_{2u}$ state then has highest eigenvalue in the triplet channel. 

{\it Summary.}---In this paper we have shown that within the
  BCS theory the leading 
instability of a phonon-mediated pairing interaction can be a triplet
state, but this must be degenerate with a singlet solution. Our
analysis relies only on the symmetries of the electron-phonon vertex
functions. We have additionally formulated a condition in terms of a
chirality operator for when this degeneracy holds. We have hence
identified the electron-phonon vertices that produce an attractive
interaction in the triplet channel and those that are
  pair-breaking, which we apply to the topological state proposed for
Cu$_x$Bi$_2$Se$_3$. If the former
dominate the latter, we show that weak electronic correlations could
stabilize the odd-parity state. Large-scale (and quantitatively
accurate) first principles calculations can in principle determine
whether specific systems (e.g. Cu$_x$Bi$_2$Se$_3$,
Sn$_{1-x}$In$_{x}$Te, etc.) satisfy the 
necessary theoretical constraints derived in our work, providing a
route to the realization of topological superconductivity in
ordinary electronic materials.

{\it Acknowledgments.}---The authors thank J. Bauer for useful
discussions. This work is supported by JQI-NSF-PFC and Microsoft Q.

\end{document}